# Integrating Multivariate Adaptive Regression Splines into Small Area Estimation for Nonlinear Poverty Modeling in Java Island

**Ajriansyah[a*], Bambang Widjanarko Otok[a], Sutikno[a]**

[a]*Department of Statistics, Institut Teknologi Sepuluh Nopember, 60111, Surabaya, Indonesia*

*Corresponding Author: jianbima1@gmail.com

**Abstract**

Small area estimation (SAE) modeling is often used to address the unreliability of estimates in areas with limited survey samples. However, standard SAE models (Fay-Herriot) can only accommodate linear patterns, whereas in real cases, including poverty, many patterns are nonlinear. This study proposes the implementation of a supervised learning model, multivariate adaptive regression splines (MARS), integrated within the SAE framework (MARS-SAE) to address the challenge of unreliable direct estimation, particularly in nonlinear cases. The hyperparameter tuning results yielded the best model with the following hyperparameter combinations: maximum basis function (Max BF) = 10, maximum interaction (MI) = 2, minimum observation (MO) = 1, penalty = 3, with general cross validation (GCV) = 6.128. MARS-SAE demonstrated very strong performance; empirically, the model proved more efficient at reducing estimation error due to its higher relative efficiency (RE) compared to the SAE Fay-Herriot model, as well as better reduction of relative standard error (RSE) than the comparison model, and it yielded a lower relative root mean squared error (RRMSE) compared to the SAE Fay-Herriot model (13.64% vs. 13.75%). MARS-SAE has been proven to capture nonlinear patterns very well while still maintaining interpretability, making it an effective tool for the predictive and diagnostic analyses of phenomena. Thus, this research can serve as an important reference for readers regionally and in Southeast Asia, given that Java Island is a strategic region for Indonesia, the largest economic power in the ASEAN region.

## 1. Introduction

The availability of accurate data has become a critically important requirement for decision-making. In the process of policy formulation and decision-making, complete data at the population level alone are not sufficient; instead, precision of information is also needed at smaller and more detailed subpopulation levels, including at the Regency, District, and even village/subdistrict levels (Chandra et al., 2018; Marchetti et al., 2025; Maulana et al., 2014; Permatasari & Ubaidillah, 2025). However, at these various subpopulation levels, sample availability is limited because national survey projects generally focus on ensuring high-quality sample selection for large populations or subpopulations, such as provinces, while at the micro level, such as regencies down to subdistricts, data quality tends to decline (Trihandika et al., 2024; Yilema et al., 2025). If the parameter estimation process in subpopulations with limited samples uses direct estimation, the results obtained will tend to have high variance and bias, making them unreliable as a basis for policy determination (Das et al., 2025; Jin et al., 2024; Marchetti et al., 2025; Sriliana et al., 2017).

Increasing the sample size can deplete resources (Gartina & Khikmah, 2020; Priatmadani et al., 2024; Trihandika et al., 2024). Thus, this issue can be addressed with Small Area Estimation (SAE), a model-based estimation method specifically designed to overcome estimation problems in areas with few or no samples at all (Dedianto & Wulansari, 2018; Fay & Herriot, 1979; Priatmadani et al., 2024). The standard SAE model, namely the SAE Fay-Herriot model, assumes a linear relationship between subpopulation parameters and auxiliary variables (Fay & Herriot, 1979; Maulana et al., 2014), which can be a drawback when confronted with nonlinear cases. Meanwhile, many real-world cases exhibit nonlinear patterns (Hosseini et al., 2024; Jin et al., 2024; Rutten et al., 2025), one of which is poverty (Amin Megat Ali et al., 2024; Xu & Xu, 2025).

Starting from the issue of the lack of reliability in direct estimation methods for estimating parameters in subpopulations with typically small sample sizes, compounded by a nonlinear relationship between the parameter of interest and its auxiliary variables, this study proposes integrating the supervised learning model multivariate adaptive regression splines (MARS) into the small area estimation (SAE) model. This resulted in the MARS-SAE model, which can estimate subpopulation parameters for nonlinear cases with high accuracy and improve estimation precision. The MARS model is chosen for several reasons: in previous research contexts, according to Hosseini (Hosseini et al., 2024), the MARS model is superior in the process of automatic knot selection compared to the penalized spline SAE model (SAE P-Spline) by Sriliana (Sriliana et al., 2017), which determines knots manually. Furthermore, compared to the Nadaraya-Watson kernel regression SAE model by Maulana (Maulana et al., 2014), MARS is better at accommodating high-dimensional cases. In addition, aside from the model's excellent performance for nonlinear cases, MARS is a nonparametric model that does not require specific data distribution assumptions and is a “white box” type of machine learning, making the modeling process and resulting outputs easily interpretable compared to “black box” machine learning models such as neural networks, making it a good model for diagnostic analysis of poverty cases and providing a deeper explanation of the phenomenon.

Java Island is a vital point for Indonesia's economy (Rahmi et al., 2025), and Indonesia is the largest economic power in terms of Gross Domestic Product (GDP) among Southeast Asian countries. Thus, Java indirectly holds a strong position in representing the regional socio-economic dynamics of Southeast Asia. Java can become one of the key regions in showcasing economic growth within the Association of Southeast Asian Nations (ASEAN), making high-quality statistics to accurately describe the condition of Java essential. Therefore, studying poverty on Java Island is important for policymaking in Indonesia and may serve as an interesting point of reference for other ASEAN countries. According to the latest publication from the Central Statistics Agency of the Republic of Indonesia (Badan Pusat Statistik/BPS), the poor population on Java Island is 12.32 million people. This figure is quite high compared to the national poor population of 23.36 million, meaning that approximately 52.7% of Indonesia's poor population lives on Java Island[1]. Based on this data, central and regional governments need to play significant roles, including formulating various policies to support poverty alleviation, which aligns with the national development plan toward Indonesia Emas 2045[2] and the international Sustainable Development Goals (SDGs), especially the first goal[3] (Das et al., 2025). Targeted policies must begin with the accurate availability of macro-level, and especially micro-level, field data, and policymakers need to have a deep understanding of the phenomena at hand. Therefore, the proposed research is highly important and is expected to have a significant impact both nationally and on a broader scale, such as for ASEAN or the Southeast Asia region.

## 2. Material and Methods

### 2.1 Small Area Estimation (SAE)

Small area estimation (SAE) is a model-based method first developed by Fay and Herriot (Fay & Herriot, 1979) and used to estimate subpopulation parameters indirectly. It was proposed to address weaknesses arising from direct estimation in subpopulations with very limited samples (Dedianto & Wulansari, 2018; Fay & Herriot, 1979; Gartina & Khikmah, 2020; Maulana et al., 2014; Priatmadani et al., 2024). The SAE model aims to improve the estimation results from the direct estimation method, which has a high bias. This improvement uses a "shrinkage factor" weighting, which works by determining the tendency of the estimate to lie closer to the direct or model estimator. The shrinkage factor relies on the ability of the SAE model to gather information ("borrowing strength") from two sources: auxiliary variables and information from other areas (Hosseini et al., 2024; Trihandika et al., 2024; Yilema et al., 2025). The basic framework of the SAE model is the linear mixed model (LLM), as shown in Equation (1) (Marchetti et al., 2025; Pusponegoro & Rachmawati, 2018).

[1] BPS RI. (2025b). Jumlah penduduk miskin. Retrieved from https://www.bps.go.id/id/query-builder. Accessed February 27, 2026

[2] Government of the Republic of Indonesia. (2024). Undang-Undang Nomor 59 Tahun 2024 tentang Rencana Pembangunan Jangka Panjang Nasional Tahun 2025-2045 [Law Number 59 of 2024 concerning the National Long-Term Development Plan for 2025-2045].

[3] United Nations General Assembly. (2015). Transforming our world: the 2030 Agenda for Sustainable Development.

$$\eta_i = \boldsymbol{x}_i^T \boldsymbol{\beta} + Z_i u_i + \varepsilon_i \tag{1}$$

where $\eta_i$ = i-th response (direct estimation), $\boldsymbol{x}_i^T$ = auxiliary variable vector, $\boldsymbol{\beta}$ = regression coefficient vector of size $p \times 1$, $Z_i$ = a known positive constant (usually 1), $u_i$ = random effect, $\varepsilon_i$ = error.

The most basic model in SAE is the Fay-Herriot model, with its optimal estimator being the empirical best linear unbiased predictor (EBLUP) (Marchetti et al., 2025; Permatasari & Ubaidillah, 2025). The SAE Fay-Herriot model is presented in Equation (2).

$$\hat{\eta}_i = \boldsymbol{x}_i^T \boldsymbol{\beta} + Z_i u_i + e_i \tag{2}$$

The model form in Equation (2) is constructed based on two elements. The first element is the sampling model.

$$\hat{\eta}_i = \eta_i + e_i \tag{3}$$

where, $i = 1, 2, \dots, n$ and $e_i$ = sampling error for the i-th area, $e_i \sim N(0, \sigma_i^2)$, where $\sigma_i^2$ = sampling variance. The second element is the linking model.

$$\eta_i = \boldsymbol{x}_i^T \boldsymbol{\beta} + Z_i u_i \tag{4}$$

where, $\boldsymbol{x}_i^T$ is an auxiliary variable vector $(1, x_{1i}, x_{2i}, \dots, x_{pi})$, $\boldsymbol{\beta}$ is the regression coefficient vector $(\beta_0, \beta_1, \dots, \beta_p)^T$, $Z_i$ is a known positive constant (usually valued at 1) and $u_i$ is the random effect for the i-th region, $u_i \sim N(0, \sigma_u^2)$, where $\sigma_u^2$ = random effects variance.

The estimated value for each region was obtained using the EBLUP estimator as in equation (5) with the aid of the shrinkage factor $(\hat{\gamma}_i)$ in equation (6).

$$\hat{\eta}_i^{EBLUP} = \hat{\gamma}_i \hat{\eta}_i + (1 - \hat{\gamma}_i) \boldsymbol{x}_i^T \widehat{\boldsymbol{\beta}} \tag{5}$$

$$\hat{\gamma}_i = \frac{\hat{\sigma}_u^2}{\hat{\sigma}_u^2 + \sigma_i^2} \tag{6}$$

## 2.2 Multivariate Adaptive Regression Splines (MARS)

Multivariate Adaptive Regression Splines (MARS) is a type of supervised learning regression model that is nonparametric (Lai et al., 2024; Seno et al., 2024), first introduced by Friedman (Friedman, 1991). MARS can handle data with complex nonlinear patterns owing to its flexible and data-driven nature, and it is reliable for high-dimensional cases (Bekar Adiguzel & Cengiz, 2023; López & Kholodilin, 2023; Nayeem et al., 2025). Furthermore, MARS can recognize patterns even when the data have unrecognized distributional forms; therefore, initial assumptions regarding the functional form of the data are not required (Friedman, 1991; López & Kholodilin, 2023; Seno et al., 2024).

The working principle of the MARS model is by partitioning each predictor variable ($x_j$) with knot points ($k_{qm}$) then there is a pair of basis functions (BF) that work in an opposite manner for each knot point. The BF pairs are expressed in Equations (7) and (8) (Sriningsih et al., 2023):

$$\begin{aligned}\left(s_{qm}\left(x_{iv(q,m)}-k_{qm}\right)\right)_+ &= \left(+\left(x_{iv(q,m)}-k_{qm}\right)\right)_+ \\ &= \max\left(0, x_{iv(q,m)}-k_{qm}\right) \\ &= \begin{cases} x_{iv(q,m)}-k_{qm}, & if\ x_{iv(q,m)} > k_{qm} \\ 0, & if\ others \end{cases}\end{aligned} \tag{7}$$

$$\begin{aligned}\left(s_{qm}\left(x_{iv(q,m)}-k_{qm}\right)\right)_+ &= \left(-\left(x_{iv(q,m)}-k_{qm}\right)\right)_+ \\ &= \max\left(0, k_{qm}-x_{iv(q,m)}\right) \\ &= \begin{cases} k_{qm}-x_{iv(q,m)}, & if\ x_{iv(q,m)} < k_{qm} \\ 0, & if\ others \end{cases}\end{aligned} \tag{8}$$

where $s_{qm}$ = basis function notation, $s_{qm} \in \{+1,-1\}$; $x_{iv(q,m)} \in \{x_j\}_{j=1}^{p}$ = predictor variable number-$v$; $k_{qm} \in \{x_{iv(q,m)}\}_{i=1}^{n}$ = knot point. Each possible knot point and its associated BF pair are candidates for constructing a MARS model, after which a greedy search strategy is used to find the BF pair that most reduces the sum of squared error (SSE).

The selection and addition of BF pairs to the MARS model were performed iteratively and automatically in the forward stage. The forward stage stops when certain criteria are met, for example, when the maximum number of BFs is reached, or the degree of interaction between predictors reaches its limit. The forward stage produces an overfitted model because it is too complex and highly adaptive to data. The overfitting problem in the forward stage is addressed during the backward stage, which involves the iterative elimination of BFs one by one, alternately removing the BF with the smallest impact on the model, based on the criterion that the smallest general cross validation (GCV) value is obtained when that BF is removed. The backward iteration stops when the pruning process results in the simplest possible model and yields a MARS model with the smallest GCV value (the optimal model) (Friedman, 1991; López & Kholodilin, 2023; Sriningsih et al., 2023).

Conceptually, the MARS model is constructed by treating basis functions (BF) as predictors for the response variable, such that the model structure appears as a linear combination of a set of BF optimally selected to regress the response. The general form of the MARS model is presented in Equations (9) – (12).

$$\eta_i = f(\boldsymbol{x}_i) + \varepsilon_i, \qquad i = 1,2,\dots,n \tag{9}$$

$$f(\boldsymbol{x}_i) = \beta_0 + \sum_{m=1}^{M} \beta_m \prod_{q=1}^{Q_m} \left(s_{qm}\left(x_{iv(q,m)} - k_{qm}\right)\right)_+ \tag{10}$$

$$\eta_i = \beta_0 + \sum_{m=1}^{M} \beta_m \prod_{q=1}^{Q_m} \left(s_{qm}\left(x_{iv(q,m)} - k_{qm}\right)\right)_+ + \varepsilon_i \tag{11}$$

$$\eta_i = \beta_0 + \sum_{m=1}^{M} \beta_m B_m(\boldsymbol{x_i}) + \varepsilon_i \tag{12}$$

where $\eta_i$ is the response variable of region-$i$, $\beta_0$ is the intercept, $\beta_m$ is the coefficient of the m-th basis function, $B_m(\boldsymbol{x_i})$ is the m-th basis function, $\boldsymbol{x}_i$ is a predictor variable vector, and $\varepsilon_i$ = random error of the i-th observation, $\varepsilon_i \sim iid\ N(0, \sigma^2)$. Therefore, in simple terms, the MARS method obtains an optimal model by running two algorithms: forward and backward. The forward iteration identifies and adds pairs of basis functions that reduce the error using a greedy search strategy; however, this can lead to overfitting. Meanwhile, the backward iteration works by pruning basis functions one by one, which introduces noise into the model, to obtain the optimal model using the smallest general cross validation (GCV) criterion (Friedman, 1991).

The MARS model used to obtain reliable estimates and integrated with SAE must employ the best models. As a machine learning model, the optimal MARS model was obtained through hyperparameter tuning. Several hyperparameters that need to be set include the maximum number of basis functions (Max BF), using the rule of 2–4 times the number of predictor variables; then maximum interaction (MI) with three options, namely 1, 2, and 3; and finally minimum observation (MO), set to 0, 1, 2, 3, and 5. Therefore, a total of combinations of models were constructed. The penalty value for GCV was 3. The best model was one of those 45 model combinations with the smallest GCV value.

### 2.3 Integration of MARS into SAE (MARS-SAE)

A MARS-SAE model was proposed in this study. In this section, the formation of the model is presented by integrating the nonlinear MARS equation into the SAE model, along with the technique for obtaining the MSE and modifying the estimator. The MARS-SAE model works by first obtaining an equation capable of handling the complex nonlinear patterns in the poverty dataset $\left(f(\boldsymbol{x}_i)\right)$ using MARS modeling, after which the nonlinear equation is used to substitute the linear equation in the SAE Fay-Herriot model $\left(\boldsymbol{x}_i^T\boldsymbol{\beta}\right)$, so that an SAE model capable of performing well on nonlinear pattern cases will be formed. In this study, the MARS-SAE modeling workflow is divided into two stages: the first stage is to obtain the best MARS model and extract the equation $\hat{f}(\boldsymbol{x}_i)$ from that model, the second stage is then to perform MARS-SAE modeling, modifying the linear SAE model by replacing its linear elements with nonlinear elements from the best MARS model identified in the first stage. Substitution of linear equations $\left(\boldsymbol{x}_i^T\boldsymbol{\beta}\right)$ in equation (2) by the nonlinear MARS equation $\hat{f}(\boldsymbol{x}_i)$ in equation (9) forms equation (13).

$$\hat{\eta}_i = \hat{f}(\boldsymbol{x}_i) + Z_i u_i + e_i \tag{13}$$

Next, the estimator used also replaces the linear elements in Equation (5) with nonlinear elements, as in Equation (14).

$$\hat{\eta}_i^{MARS-SAE} = \hat{\gamma}_i\hat{\eta}_i + (1-\hat{\gamma}_i)\hat{f}(\boldsymbol{x}_i) \tag{14}$$

Meanwhile, the shrinkage factor is calculated using Equation (6) is still used.

In contrast to the SAE Fay-Herriot EBLUP model, whose model MSE can be derived analytically, the MARS-SAE model is a nonparametric model that does not have a closed-form MSE because of its adaptively flexible nature. Based on this, this study employs the bootstrap method to obtain the MSE of the MARS-SAE model. The type of bootstrap used is parametric bootstrap with B = 10000 replications; additionally, sampling variance is retained, and the uncertainty of the MARS model is incorporated. MSE serves as the basic element used to estimate the evaluation metrics for the model, where the main evaluation metric used is the relative root mean squared error (RRMSE). The performance of the proposed model is compared with that of the Fay-Herriot model.

## 2.4 Dataset

### 2.4.1 Data Sources

The dataset used in this study was sourced from the published data of the Central Statistics Agency (BPS) of the Republic of Indonesia[4]. The research object is the result of the direct estimation of the percentage of the poor population in each Regency/City on the island of Java, totaling 119 regions in 2024. The auxiliary variables used include the percentage of households with access to proper sanitation, percentage of households

[4] BPS RI. (2026). Badan Pusat Statistik Republik Indonesia Produk-Tabel Dinamis. Retrieved from https://www.bps.go.id/id/query-builder. Accessed January 11, 2026

with access to proper drinking water, labor force participation rate, percentage of per capita expenditure for food, and percentage of the population with health insurance.

### 2.4.2 Research Variables

As previously explained, this study involves five auxiliary variables in estimating the percentage of the poor population. Further details on the research variables are presented in Table 1.

Table 1. Research Variables

| Variables | Scale (Unit) |
|---|---|
| Percentage of Poor Population (Y) | Ratio (%) |
| Percentage of Households with Adequate Sanitation (X1) | Ratio (%) |
| Percentage of Households with Access to Safe Drinking Water (X2) | Ratio (%) |
| Labor Force Participation Rate (X3) | Ratio (%) |
| Percentage of Per Capita Expenditure on Food (X4) | Ratio (%) |
| Percentage of Population with Health Insurance (X5) | Ratio (%) |

The nonlinear relationship between the response and predictor variables is shown in Figure 1.

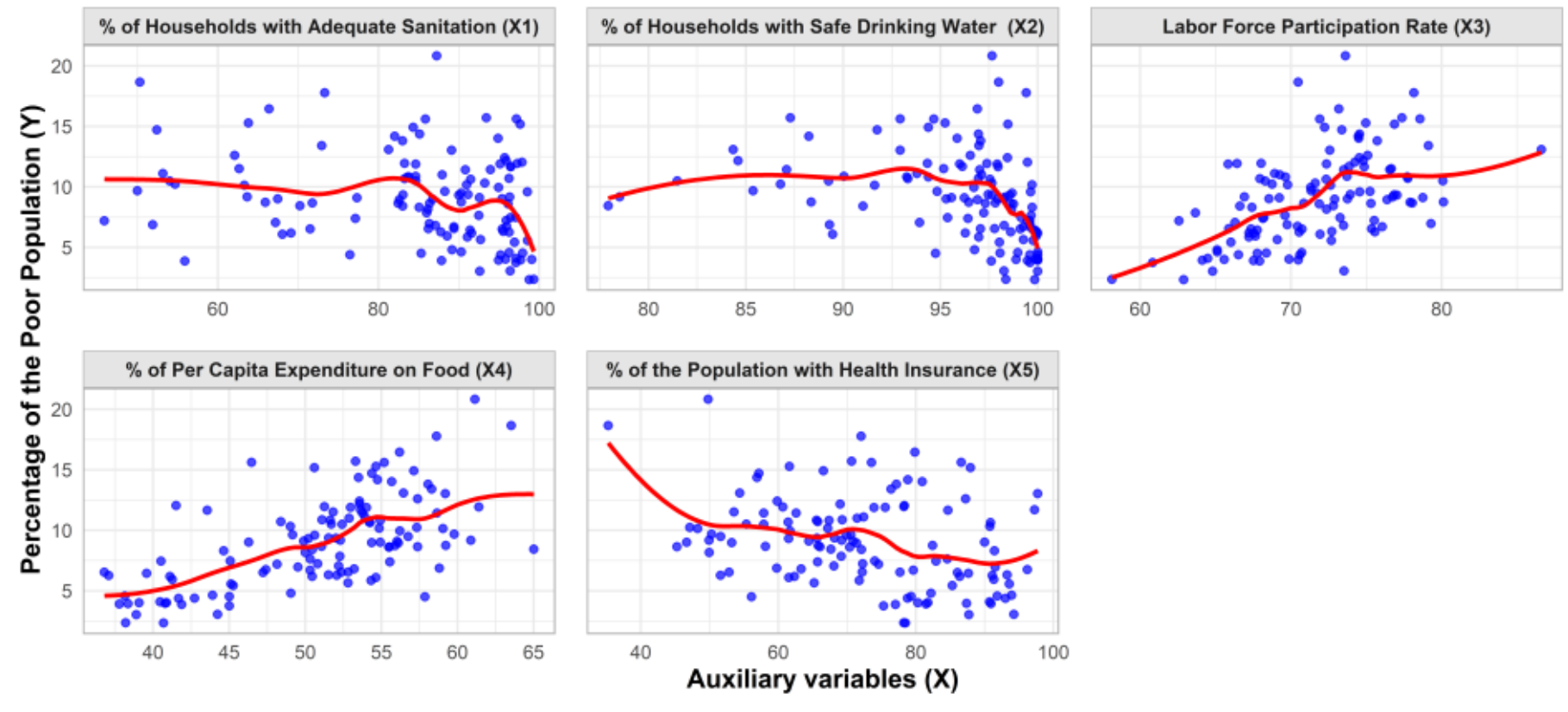

Figure 1. Pattern of the Relationship between Poverty Percentage and Auxiliary Variables

Based on Figure 1, the relationship between the five predictor variables and the response tends to form a nonlinear pattern, with the variable representing the percentage of households with access to improved drinking water being the most nonlinear.

2.4.3 Sampling Variance and Relative Standard Error

The direct estimation values for the percentage of the poor population for 119 regions in Java Island are secondary data from BPS, in which the dataset does not provide sampling variance information for all regencies/cities, while this information is essential for SAE modeling. Out of a total of 119 Regencies/Cities spread across 6 Provinces on Java Island (East Java, Central Java, West Java, DI Yogyakarta, Banten, and DKI Jakarta), only 78 Regencies/Cities spread across 3 Provinces, namely East Java, Central Java, and DI Yogyakarta, have a complete dataset with Relative Standard Error (RSE) values. The other 41 regencies/cities do not have RSE. One solution that has been implemented is to model using the available RSE to estimate RSE values for regions with incomplete data. With the availability of the RSE, the sampling variance is calculated using Equation 15 below.

$$\sigma_i^2 = \left(SE(\hat{\eta}_i)\right)^2 = \left(\frac{RSE(\hat{\eta}_i)}{100\%} \times \hat{\eta}_i\right)^2 \tag{15}$$

where $\sigma_i^2$ = sampling variance, and $SE$ = standard error.

The practice of obtaining estimates of sampling variance through modeling is reasonable when information on the sampling variance of survey results is not fully available, as supported by research by Das (Das et al., 2025), Sugasawa (Sugasawa & Kubokawa, 2020) and You (You & Hidiroglou, 2023).

Based on the BPS framework, published estimated values for the public are divided into two categories according to their RSE values: if a region has an RSE value < 25%, then the direct estimation results for subpopulation parameters in that region are considered accurate for policy-making purposes. If the RSE is in the range of 25% ≤ RSE ≤ 50%, it may still be used, but with caution and supplemented by additional information when serving as a decision-making source. The application of the SAE model is highly recommended, especially to address shortcomings in situations where the RSE is greater than or equal to 25% but still less than or equal to 50%[5]. In the case examined in this study, several regions still have actual RSE (not model-based estimates) that fall into the 25% ≤ RSE ≤ 50% category.

[5] BPS RI. (2025a). Governance and Framework for Small Area Estimation.

## 3. Results

### 3.1 Exploratory Data Analysis (EDA)

EDA is a crucial step before modeling. In this analysis, the dataset will be described both descriptively and diagnostically as an in-depth examination of each case. Table 2 presents the descriptive statistics of the data.

Table 2. Descriptive Statistics

| Statistics | Variables | | | | | |
|---|---|---|---|---|---|---|
| | Y | X1 | X2 | X3 | X4 | X5 |
| Minimum | 2.34 | 45.88 | 77.92 | 58.13 | 36.80 | 35.25 |
| 1st Quartile | 6.38 | 82.23 | 95.00 | 67.99 | 46.38 | 62.03 |
| Median | 8.98 | 88.55 | 97.55 | 71.92 | 52.19 | 72.05 |
| Mean | 9.07 | 84.75 | 96.05 | 71.58 | 50.88 | 72.33 |
| 3rd Quartile | 11.42 | 95.45 | 99.21 | 74.83 | 55.34 | 82.67 |
| Maximum | 20.83 | 99.34 | 100.00 | 86.62 | 65.01 | 97.71 |

Based on Table 2, the response variables have a high level of heterogeneity. There is an area with a very high percentage of poor population at 20.83%, with an average of approximately 9%. The substantial gap between the minimum and maximum values indicates a high level of economic disparity among the regencies/municipalities on the island of Java. Meanwhile, among the auxiliary variables, the percentage of the population with health insurance is the most heterogeneous.

Next, visualization was performed to measure the strength of the relationship between each auxiliary variable and the percentage of the poor population. The Pearson correlation test was used in this section and is visualized in Figure 2.

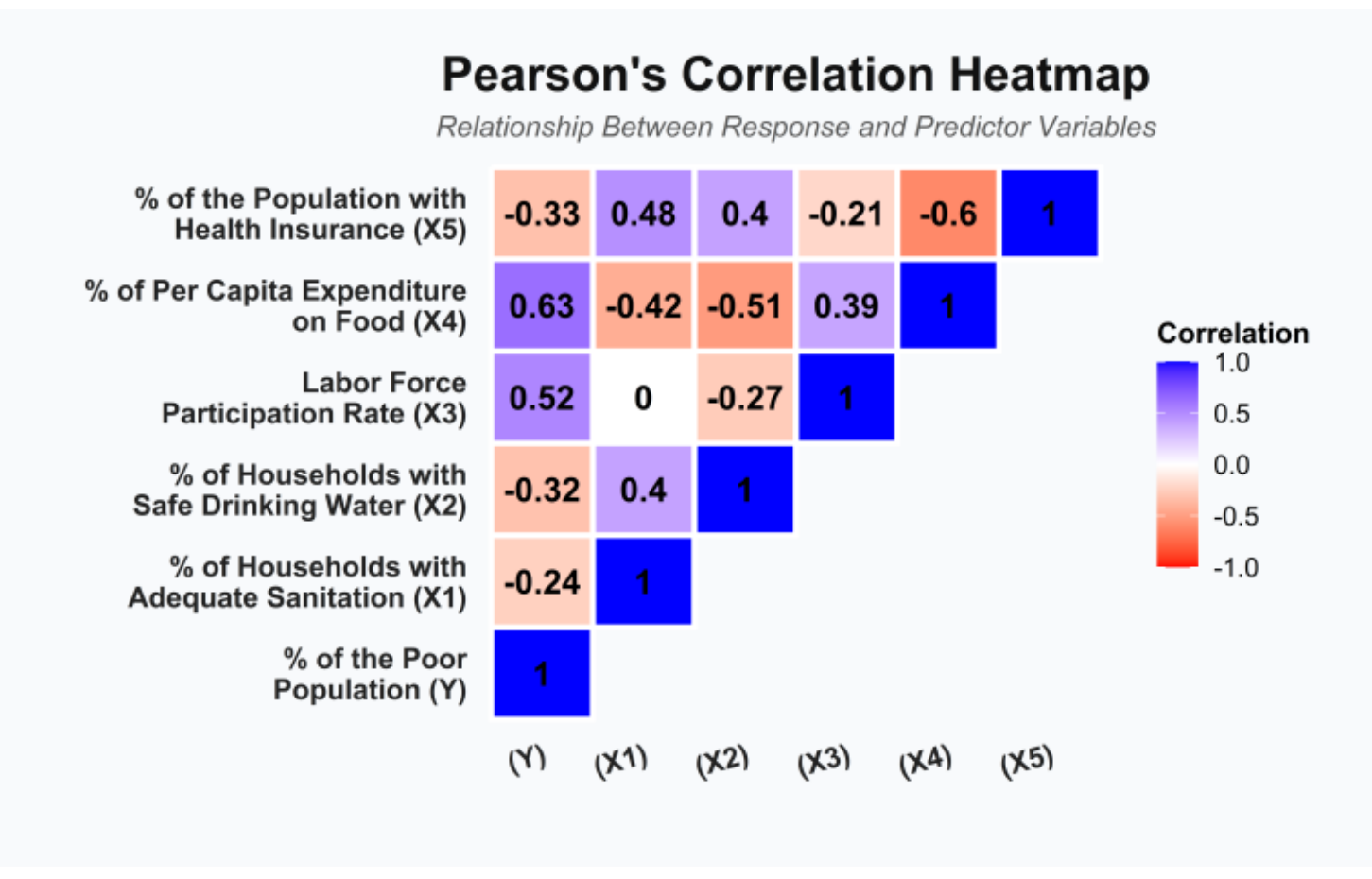

Figure 2. Strength of Correlation between Auxiliary Variables and Response (Pearson's correlation)

Figure 2 shows the strength of the correlation between the percentage of the poor population and each supporting variable, all of which are significantly correlated with the poverty rate. The variable with the strongest correlation was the percentage of per capita expenditure on food. The direction of the correlation is positive, which is consistent with Engel's theory that as income increases, the proportion of expenditure on food decreases (Engel, 2021). In other words, regions with a high percentage of poor people also have a high percentage of per capita expenditure on food.

### 3.2 SAE Fay-Herriot Model

The SAE Fay-Herriot modeling with the EBLUP estimator yields the model presented in equation (16).

$$\begin{aligned}\hat{\eta}_i = -28.30 &- 0.02(\text{\% of Households with Adequate Sanitation})_i \\ &+ 0.03(\text{\% of Households with Safe Drinking Water})_i \\ &+ 0.27(\text{Labor Force Participation Rate})_i \\ &+ 0.29(\text{\% of Per Capita Expenditure on Food})_i \\ &+ 0.02(\text{\% of the Population with Health Insurance})_i \\ &+ Z_i u_i + e_i\end{aligned} \tag{16}$$

In the SAE Fay-Herriot model, only the labor force participation rate and the percentage of per capita expenditure on food contributed significantly. The primary evaluation metric for measuring model performance was the RRMSE for both the SAE Fay-Herriot model and the proposed MARS-SAE model. The SAE Fay-Herriot EBLUP model produced an RRMSE of 13.75%.

### 3.3 Multivariate Adaptive Regression Splines Model

Hyperparameter tuning was first conducted to obtain the optimal model combination (the best MARS model). A total of 45 model scenarios consisting of hyperparameter combinations: Max BF = (10, 15, 20); MI = (1, 2, 3); MO = (0, 1, 2, 3, 5); and penalty = 3 were used in this study. After modeling, the hyperparameter combination that produced the best MARS model was Max BF = 10, MI = 2, MO = 1, and penalty = 3, with a GCV of 6.128. The nonlinear equation $\hat{f}(\boldsymbol{x}_i)$ from the best MARS model can be written in equation (17).

$$\hat{f}(\boldsymbol{x}_i) = 12.07 - 0.71(BF_1) - 0.40(BF_2) + 0.33(BF_1 BF_3) \tag{17}$$

The form of each BF function is presented in Table 3.

Table 3. Functional Forms of BFs

| Basis Functions |
|---|
| $BF_1 = h(X2 - 94.33)$ |
| $BF_2 = h(74.5 - X3)$ |
| $BF_3 = h(X4 - 54.25)$ |

The results of the MARS modeling were further used in MARS-SAE modeling.

### 3.4 MARS-SAE Model

MARS-SAE modeling occurs in two stages, where the first stage is obtaining the best MARS model and then using $\hat{f}(\boldsymbol{x}_i)$ from the model as the model estimator and then the second stage is to generate estimates of the percentage of poor population using the estimator that has been modified for MARS-SAE after using the shrinkage factor that serves as a bridge between the direct estimator ($\hat{\eta}_i$) with the model estimator $\hat{f}(\boldsymbol{x}_i)$. One of the main objectives of the SAE is to increase estimation precision by reducing the standard error (SE) in each observed area. The average RSE reduction percentage was 15.60%, with an RRMSE of 13.64%. The importance levels of the auxiliary variables are presented in Table 4.

Table 4. Importance Level of Auxiliary Variables

| Variables | Importance | |
|---|---|---|
| | GCV | RSS |
| % of Per Capita Expenditure on Food (X4) | 100.00 | 100.00 |
| % of Households with Safe Drinking Water (X2) | 61.50 | 62.30 |
| Labor Force Participation Rate (X3) | 61.50 | 62.30 |
| % of Households with Adequate Sanitation (X1)-unused | 0.00 | 0.00 |
| % of the Population with Health Insurance (X5)-unused | 0.00 | 0.00 |

Based on Table 4, the most dominant variable in this model is the percentage of per capita expenditure on food, followed by the percentage of households with access to improved drinking water and the labor force participation rate. Meanwhile, the variables that are not important (not used) in the model are the percentage of households with improved sanitation and the percentage of the population with health insurance coverage.

The distribution of the estimated percentages of the poor population in each Regency/City on Java Island using the MARS-SAE model is visualized in the form of a thematic map presented in Figure 3.

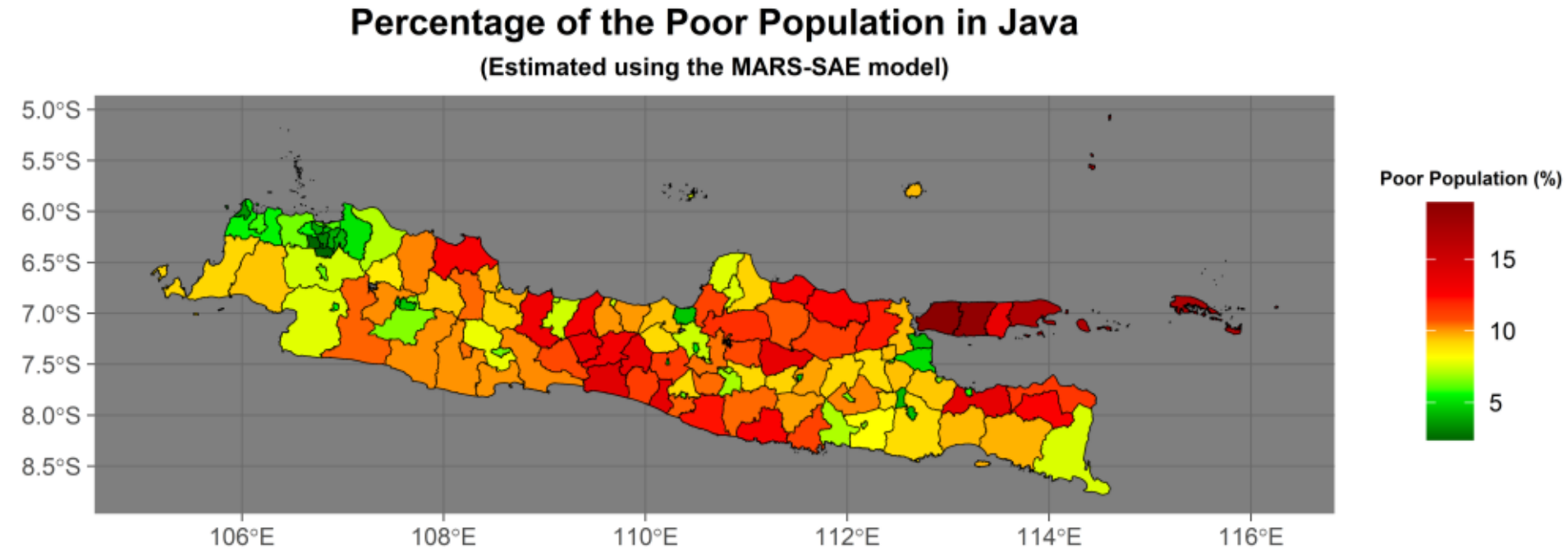

Figure 3. Thematic Map of Estimated Percentage of Poor Population using MARS-SAE

In Figure 3, economic disparity can be seen, where most regencies/cities in Java Island have a poverty rate above 8.25% (the national average). A total of 57.98% of the 119 regions have a poverty rate above the national average; at the highest point is Bangkalan Regency, reaching 18.94%, followed by Sampang Regency at 18.51%, and Sumenep Regency with a poverty rate of 16.80%. These three regions are neighboring areas located in the Madura Island region. A vastly different situation is found in major cities such as Surabaya and Jakarta, with poverty rates of approximately 3% to 6%. The high poverty levels in the Madura Island region indicate a spatial effect; similar patterns also appear in adjacent areas in Central and West Java and the areas surrounding Jakarta.

### 3.5 Model Performance Comparison

The models used for estimation will be evaluated for their performance, and each will be compared with the others to assess the effectiveness of the proposed model. In addition to the RRMSE as the primary evaluation metric, several other measures were used to compare the model performance. The metrics are relative efficiency (RE), percentage reduction in relative standard error (RSE), and R-squared. The evaluation metrics for each model are listed in Table 5.

Table 5. Comparison of Evaluation Metrics for Each Model

| Metrics | Models | |
|---|---|---|
| | SAE Fay-Herriot | MARS-SAE |
| RRMSE (%) | 13.75 | 13.64 |
| Max RRMSE (%) | 25.78 | 23.02 |
| RE | 1.454 | 1.466 |
| RSE Reduction (%) | 14.47 | 15.60 |
| R-Squared (%) | 93.37 | 93.59 |

To clarify the comparative positions of each model, the descriptive statistics of the direct RSE estimation for regencies/cities with available real data can be presented, where the actual RSE for the percentage of poor population cases on Java Island has the following descriptive statistics (in %): mean = 16.47; minimum = 10.15; 1st quartile = 12.65; median = 14.76; 3rd quartile = 19.93; and maximum = 29.29. Thus, because the RRMSE and RSE values can be directly compared, overall, both models have RRMSE values that are lower than the average direct estimation RSE. MARS-SAE has the lowest RRMSE value, and, more interestingly, MARS-SAE can produce an RRMSE with a maximum value of 23.02%, meaning every estimate produced by the proposed model has an RRMSE/RSE value of < 25%, thus falling into the accurate category. In contrast, the RRMSE/RSE values produced by the SAE Fay-Herriot model and direct estimation still include several areas with conditions of 25% ≤ RRMSE/RSE ≤ 50%, which should be used with caution for decision-making purposes.

Furthermore, in terms of the RE aspect, the MARS-SAE model is superior to the comparison model. On the RSE reduction metric, the MARS-SAE model reduces substantially more than the Fay-Herriot model. The performance of MARS-SAE excels even in R-squared, further demonstrating its ability to outperform all evaluation metrics. The overall comparison results show that the MARS-SAE model's performance is significantly superior to the Fay-Herriot SAE model. The conclusion of this comparison supports the findings of Frink (Frink & Schmid, 2025) and Hosseini (Hosseini et al., 2024).

## 4. Discussion

This study proposes a Small Area Estimation model capable of handling nonlinear cases between the area parameter being estimated and its auxiliary variables using the supervised learning model Multivariate Adaptive Regression Splines to provide estimation values with higher precision. The proposed model is called MARS-SAE, which is a theoretical development of the SAE Fay-Herriot EBLUP by applying the nonlinear equations of the MARS model to SAE, which fundamentally assumes a linear equation.

Theoretically, MARS-SAE is proposed because of its advantages, including that observed poverty cases violate the linearity assumption, meaning the nonlinear MARS model can provide accurate estimations. MARS exhibits strong adaptability to data, even without a specific distribution. MARS can automatically determine knot locations, thereby avoiding researcher subjectivity while capturing the variable interactions. MARS is a type of machine learning model that can be interpreted to deepen the diagnostic analysis of the investigated phenomena.

Empirical evidence of the superiority of MARS-SAE over the Fay-Herriot SAE is demonstrated through better model performance across all evaluation metrics. Starting from the lowest RRMSE, there was a reduction in RSE of 15.60%, which was higher than the RSE reduction achieved by the linear SAE, an RE of 1.466 indicating that the proposed MARS-SAE estimator was 0.82% more efficient than the linear SAE estimator, as well as a higher R-squared of 0.9359, meaning the model was able to explain 93.59% of the variation in the percentage of poor population, with the remainder explained by other factors outside the model. Therefore, MARS-SAE was the best model in this study.

Furthermore, an equally important discussion highlights the performance of MARS-SAE and linear SAE in obtaining MSE, in which the estimated MSE values for each area play a crucial role in reducing RSE and calculating RE. In this case, MARS-SAE can make significant improvements (decreases in MSE) in certain areas with high errors, although not uniformly across all areas, whereas linear SAE provides uniform improvements but results in less sharp MSE reductions than MARS-SAE. The identification results show that the proposed model is more efficient overall, and there are certain areas that are difficult to interpret with the linear model, indicating a strong nonlinear pattern that MARS-SAE can capture.

## 5. Conclusion

The findings of this study enrich the small area estimation approach by employing a nonlinear method through the integration of the MARS model into SAE. These results enhance the reputation of the supervised learning model MARS in modeling nonlinear cases on the topic of poverty. The success of MARS-SAE stands out because it is globally more efficient than the benchmark model, namely the SAE Fay-Herriot, as evidenced by its higher RE values. The proposed model outperformed the benchmark model across all evaluation metrics, including the RRMSE as the primary metric. Thus, this study successfully introduces an SAE model that is efficient in capturing nonlinear patterns and produces estimation results that significantly reduce errors.

The percentage of per capita expenditure on food is the most dominant variable in explaining the percentage of the poor population, making this aspect and its supporting indicators key factors in controlling the poverty rate on the island of Java. In addition, other auxiliary variables—except the variable for the percentage of households with proper sanitation and the percentage of the population with health insurance—also contribute and should be considered in policy formulation by policymakers at both regional and central government levels. This study emphasizes the importance of micro-level statistics to enable more objective decision-making that reflects the conditions on the ground. Accordingly, MARS-SAE is not only a model capable of predicting poverty figures but can also be used as an incisive diagnostic analysis tool for poverty cases under study.

## 6. Research limitations and potential for further research

This study has limitations. The actual sampling variance values for 41 of the 119 regions could not be obtained; therefore, modeling was necessary to estimate the sampling variance in regions where it was unavailable. In future studies, the use of a complete dataset with the actual sampling variance for each region is highly recommended. Additionally, there are various potential directions for future research, including incorporating a feature selection process prior to the modeling. While MARS can identify important variables, feature selection can help reduce noise in the dataset and may improve model performance. Of course, the results should be compared fairly before and after the feature selection process is applied. Another potential direction is to develop the model by adding spatial elements to better capture spatial patterns, as the mapping results indicate a spatial pattern in poverty cases on the island of Java. Future studies may also utilize “black box” machine learning models if the goal is to achieve superior accuracy without requiring an in-depth interpretation or diagnostic analysis of the case under study.

## Author's Contributions

A. and B.W.O. contributed to the conceptualization, methodology, and formal analysis of this study. A. was responsible for software use, project administration, and drafting the original manuscript. B.W.O. was responsible for validation, resource allocation, and supervision. S. was responsible for the investigation and data curation. B.W.O. and S. also contributed to the writing—review and editing—the manuscript. The authors have read and approved the final version of the manuscript.

## Declaration of Competing Interest

The authors declare that there is no conflict of interest regarding the publication of this paper.

## Funding Information

This research was funded by the Indonesian Endowment Fund for Education (LPDP) through the LPDP scholarship with the scholarship recipient identification number: 202501110000650.

## Declaration of AI Use

During the preparation of this manuscript, the author used "Paperpal by Editage" to translate the Indonesian manuscript into English and to check and correct the grammar. After using these tools, the author remained attentive, reviewed the accuracy of the results, edited the content as needed, and bears full responsibility for the entire content of the manuscript.

## Acknowledgment

Through this writing, the authors would like to express their utmost gratitude to the Ministry of Finance of the Republic of Indonesia, especially to the Indonesia Endowment Fund for Education (LPDP), for all the support that has been given to the authors through the LPDP scholarship.

The researchers would also like to express their sincere gratitude to the Institut Teknologi Sepuluh Nopember, especially the Department of Statistics, for providing the opportunity to gain knowledge and for ensuring comfort and a supportive environment in the advancement of science.

## References

Amin Megat Ali, M. S., Zabidi, A., Md Tahir, N., Mohd Yassin, I., Eskandari, F., Saadon, A., Taib, M. N., & Ridzuan, A. R. (2024). Short-term Gini coefficient estimation using nonlinear autoregressive multilayer perceptron model. *Heliyon*, *10*(4), e26438. https://doi.org/10.1016/j.heliyon.2024.e26438

Bekar Adiguzel, M., & Cengiz, M. A. (2023). Model selection in multivariate adaptive regressions splines (MARS) using alternative information criteria. *Heliyon*, *9*(9), e19964. https://doi.org/10.1016/j.heliyon.2023.e19964

Chandra, H., Salvati, N., & Chambers, R. (2018). Small area estimation under a spatially non-linear model. *Computational Statistics and Data Analysis*, *126*, 19–38. https://doi.org/10.1016/j.csda.2018.04.002

Das, S., Deepawansa, D., & Lahiri, P. (2025). Multidimensional Poverty Mapping for Small Areas. *ArXiv Preprint*.

Dedianto, D., & Wulansari, I. Y. (2018). Aplikasi Small Area Estimation (SAE) Metode Pseudo- EBLUP Pada Official Statistics Di Indonesia Studi Kasus : Estimasi Pengeluaran Rumah Tangga di Provinsi Jawa Timur Tahun 2016. *Jurnal Aplikasi Statistika & Komputasi Statistik*, *10*(2). https://doi.org/10.34123/jurnalasks.v10i2.89

Engel, E. (2021). Die vorherrschenden Gewerbszweige in den Gerichtsämtern mit Beziehung auf die Productions- und Consumtionsverhältnisse des Königreichs Sachsen. *WISTA – Wirtschaft Und Statistik*, *73*(2), 126–136. https://www.econstor.eu/handle/10419/233572 (The original work was published in 1857).

Fay, R. E., & Herriot, R. A. (1979). Estimates of Income for Small Places: An Application of James-Stein Procedures to Census Data. *Journal Ofthe American Statistical Association*, *74*(366), 269–277. https://doi.org/10.1080/01621459.1979.10482505

Friedman, J. H. (1991). Multivariate Adaptive Regression Splines. *The Annals of Statistics*, *19*(1), 1–141. https://doi.org/10.1214/aos/1176347963

Frink, N., & Schmid, T. (2025). Small area prediction of counts under machine learning-type mixed models. *Computational Statistics and Data Analysis*, *211*. https://doi.org/10.1016/j.csda.2025.108218

Gartina, D., & Khikmah, L. (2020). Pendugaan Kemiskinan Menggunakan Small Area Estimation Dengan Pendekatan Emperical Best Linier Unbiased Prediction (EBLUP). *Statistika*, *8*(2). https://doi.org/10.26714/jsunimus.8.2.2020.159-165

Hosseini, S. E., Shahsavani, D., Rabiei, M. R., & Arashi, M. (2024). Small area estimation with partially linear mixed-t model with measurement error. *Journal of Computational and Applied Mathematics*, *446*(September 2023), 115871. https://doi.org/10.1016/j.cam.2024.115871

Jin, I. H., Liu, F., Park, J., Eugenio, E., & Liu, S. (2024). Bayesian hierarchical spatial model for small-area estimation with non-ignorable nonresponses and its application to the NHANES dental caries data. *Journal of the Korean Statistical Society*, *53*(4), 949–969. https://doi.org/10.1007/s42952-024-00274-3

Lai, V. Q., Kounlavong, K., Keawsawasvong, S., Bui, T. S., & Huynh, N. T. (2024). A machine learning regression approach for predicting uplift capacity of buried pipelines in anisotropic clays. *Journal of Pipeline Science and Engineering*, *4*(1), 100147. https://doi.org/10.1016/j.jpse.2023.100147

López, F., & Kholodilin, K. (2023). Putting MARS into space. Non-linearities and spatial effects in hedonic models. *Regional Science Association International*, 871–896. https://doi.org/10.1111/pirs.12738

Marchetti, S., Salvati, N., Fabrizi, E., & Tzavidis, N. (2025). Small area estimation of equivalized income for local labour systems in Italy via M-quantile area-level models. *Statistical Methods & Applications*, *34*(3), 449–470. https://doi.org/10.1007/s10260-025-00791-3

Maulana, U., Darsyah, M. Y., & Utami, T. W. (2014). Small area estimation untuk pendugaan jumlah penduduk miskin di kota semarang dengan pendekatan kernel-bootstrap. *Statistika*, *2*(2). https://garuda.kemdiktisaintek.go.id/documents/detail/1767627

Nayeem, A., Bonakdari, H., Khalloufi, S., & Aider, M. (2025). Application of multivariate adaptive regression splines (MARS) to study the colorization occurring in the process of lactulose production following lactose electro-activation. *International Dairy Journal*, *168*, 106291. https://doi.org/10.1016/j.idairyj.2025.106291

Permatasari, N., & Ubaidillah, A. (2025). Small Area Estimation of poverty using remote sensing data. *Statistical Journal of the IAOS*, *41*(1), 180–190. https://doi.org/10.1177/18747655241308390

Priatmadani, Sari, P. P., Rahmat, E. N., Aji, P. B., Nafiis, F. A., & Istiana, N. (2024). Small Area Estimation of Maluku and Papua Island Child Poverty Levels in 2023. *Jurnal Matematika, Statistika Dan Komputasi*, *21*(1), 46–61. https://doi.org/10.20956/j.v21i1.35293

Pusponegoro, N. H., & Rachmawati, R. N. (2018). Spatial Empirical Best Linear Unbiased Prediction in Small Area Estimation of Poverty. *Procedia Computer Science*, *135*, 712–718. https://doi.org/10.1016/j.procs.2018.08.214

Rahmi, S., Sunusi, N., & Ilyas, N. (2025). Spatial autoregressive quantile regression modeling of gross regional domestic product data in Java Island. *MethodsX*, *15*. https://doi.org/10.1016/j.mex.2025.103621

Rutten, S., Sumalinab, B., Gressani, O., Neyens, T., Duarte, E., Hens, N., & Faes, C. (2025). Penalized distributed lag non-linear models for small area data using Laplacian-P-splines. *Statistics and Computing*, *36*(38). https://doi.org/10.1007/s11222-025-10790-9

Seno, M. E., Zeini, H. A., Imran, H., Noori, M., Henedy, S. N., & Ghazaly, N. M. (2024). Advancing in creep index of soil prediction: A groundbreaking machine learning approach with Multivariate Adaptive Regression Splines. *Results in Materials*, *24*, 100641. https://doi.org/10.1016/j.rinma.2024.100641

Sriliana, I., Sunandi, E., & Rafflesia, U. (2017). Pemodelan Kemiskinan di Provinsi Bengkulu Menggunakan Small Area Estimation dengan Pendekatan Semiparametrik Penalized Spline. *Jurnal MIPA*, *40*(2), 134–140.

https://garuda.kemdiktisaintek.go.id/documents/detail/541464

Sriningsih, R., Otok, B. W., & Sutikno. (2023). Determination of the best multivariate adaptive geographically weighted generalized Poisson regression splines model employing generalized cross-validation in dengue fever cases. *MethodsX*, *10*, 102174. https://doi.org/10.1016/j.mex.2023.102174

Sugasawa, S., & Kubokawa, T. (2020). Small area estimation with mixed models: a review. *Japanese Journal of Statistics and Data Science*, *3*(2), 693–720. https://doi.org/10.1007/s42081-020-00076-x

Trihandika, L. F., Ubaidillah, A., Az-Zahra, A., Amirudin, A., Rusydiana, M., & Maharani, Z. (2024). Small Area Estimation Anak Tidak Sekolah di Pulau Kalimantan Tahun 2023. *Limits: Journal of Mathematics and Its Applications*, *21*(2), 273–288. https://doi.org/10.12962/limits.v21i2.21497

Xu, B., & Xu, R. (2025). How can government expenditure effectively achieve energy poverty reduction? A non-linear perspective. *Energy*, *335*, 137760. https://doi.org/10.1016/j.energy.2025.137760

Yilema, S. A., Shiferaw, Y. A., Fenta, H. M., & Belay, A. T. (2025). Estimating the local - level child full polio vaccination rates in Ethiopia using a hierarchical Bayes small area estimation approach. *Discover Public Health*, *22*(292). https://doi.org/10.1186/s12982-025-00695-3

You, Y., & Hidiroglou, M. (2023). Application of Sampling Variance Smoothing Methods for Small Area Proportion Estimation. *Journal of Official Statistics*, *39*(4), 571–590. https://doi.org/10.2478/jos-2023-0026